\documentclass[aip,apl,preprint,numerical]{revtex4-1} 

\usepackage[utf8]{inputenc} 

\usepackage{geometry} 
\usepackage{graphicx} 
\usepackage{epstopdf}

\begin{document}

\title{Controlled epitaxial graphene growth within amorphous carbon corrals}
\author{James Palmer}
\affiliation{School of Physics, Georgia Institute of Technology, Atlanta Georgia}
\author{Jan Kunc}
\affiliation{School of Physics, Georgia Institute of Technology, Atlanta Georgia}
\affiliation{Faculty of Mathematics and Physics, Institute of Physics, 12116 Prague, Czech Republic}
\author{Yike Hu}
\author{John Hankinson}
\author{Zelei Guo}
\affiliation{School of Physics, Georgia Institute of Technology, Atlanta Georgia}
\author{Claire Berger}
\affiliation{School of Physics, Georgia Institute of Technology, Atlanta Georgia}
\affiliation{CNRS - Institut Néel, Grenoble, France}
\author{Walt de Heer}
\affiliation{School of Physics, Georgia Institute of Technology, Atlanta Georgia}
\date{\today} 

\begin{abstract}

Structured growth of high quality graphene is necessary for technological development of carbon based electronics. Specifically, control of the bunching and placement of surface steps under epitaxial graphene on SiC is an important consideration for graphene device production. We demonstrate lithographically patterned evaporated amorphous carbon corrals as a method to pin SiC surface steps. Evaporated amorphous carbon is an ideal step-flow barrier on SiC due to its chemical compatibility with graphene growth and its structural stability at high temperatures, as well as its patternability. The amorphous carbon is deposited in vacuum on SiC prior to graphene growth. In the graphene furnace at temperatures above 1200$^\circ$C, mobile SiC steps accumulate at these amorphous carbon barriers, forming an aligned step free region for graphene growth at temperatures above 1330$^\circ$C. AFM imaging and Raman spectroscopy support the formation of quality step-free graphene sheets grown on SiC with the step morphology aligned to the carbon grid.
\end{abstract}

\maketitle

\begin{figure}
\includegraphics[height=3in]{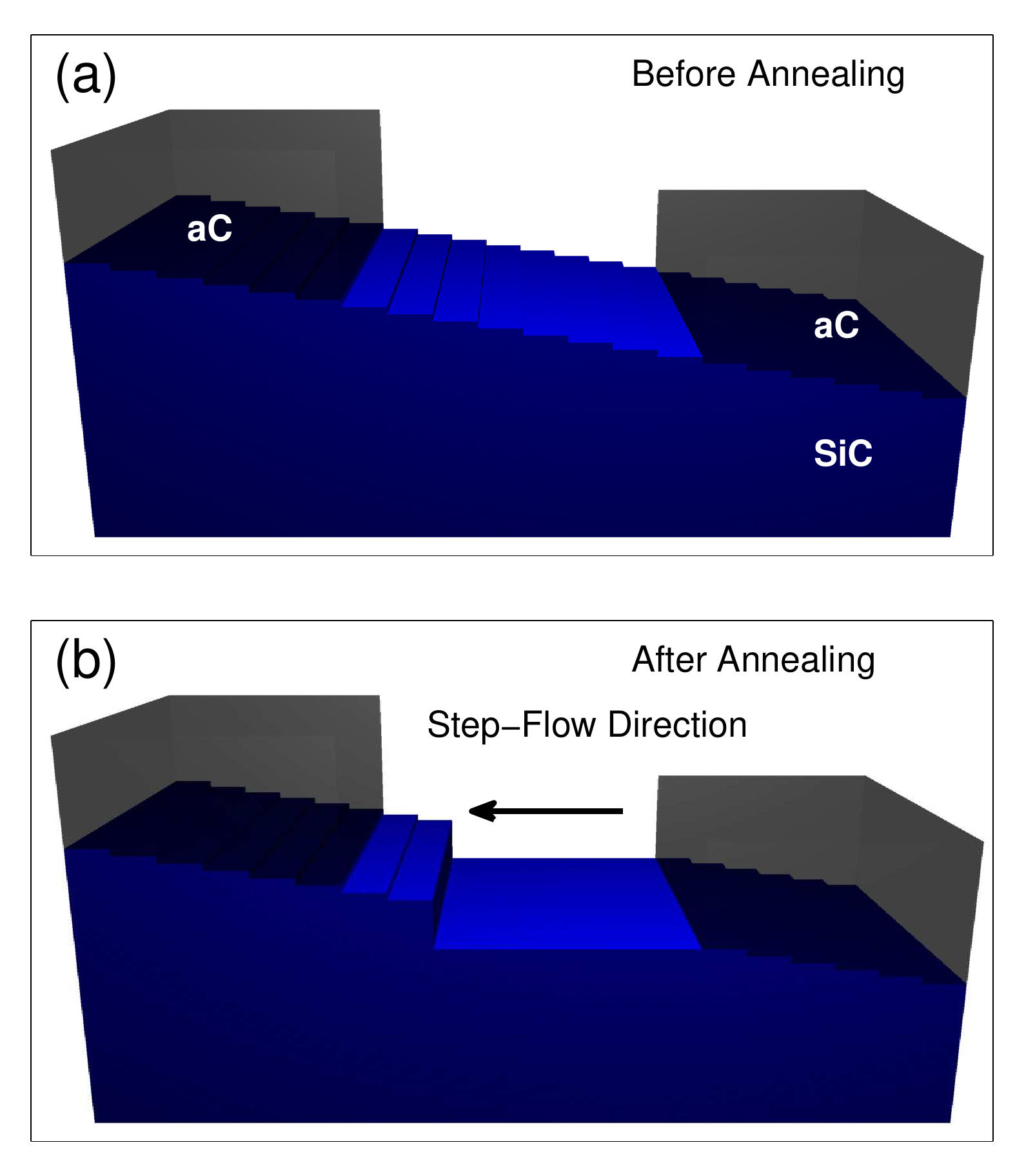}
\caption{aC step pinning process. Depicted in (a) are the SiC steps occuring before the graphene growth. In (b) the mobile steps have accumulated to the lower free energy configuration.}
\label{schematic}
\end{figure}

To be technologically viable, graphene must be grown in a well controlled manner. Beyond preparing well ordered 2-dimensional graphene films, growth needs to be tailored at the nanoscale to produce the desired devices at pre-defined locations.  Epitaxial graphene on silicon carbide (EG)  meets these criteria\cite{jphysb2004}. EG has excellent electronic transport properties, including record high frequency operation \cite{guo2013},  long spin diffusion lengths\cite{dlubak_highly_2012}, and room temperature ballistic transport\cite{ballistic}. Precise EG growth has been successful on tailored SiC substrates. Single layer nanoribbon growth has been produced on the SiC sidewalls of plasma etch steps \cite{sprinkle, ballistic} and EG thickness control was demonstrated on etched mesas \cite{handa}.  Selective graphene growth was also achieved by capping flat SiC with a high temperature stable layer to prevent (or enhance) graphene formation\cite{roy10, puybaret, camara}. Progress in graphene growth selectivity was shown using Chemical Vapor Deposited (CVD) methods, where an oxide barrier\cite{ADMA:ADMA201104195} provides better growth control on Cu and self-assembled nanoribbons on gold surfaces have been demonstrated.\cite{caiatomically2010} Unlike CVD methods, however, the advantage of EG for technological development is to be scalable directly on a high quality semi-insulating production substrate. Further advantages of EG include its integration with silicon \cite{dong} and SiC electronics\cite{weber}.

This Letter focuses on locally controlling the step bunching of SiC during EG growth. Although EG grows like an overlaying carpet on the stepped SiC surface, steps provide graphene nucleation sites \cite{Ohta, fanming, handa} that need to be controlled. Furthermore, underlying SiC steps induce additional electronic scattering in graphene. \cite{ross} SiC step spacing depends on the miscut angle of the SiC crystal surface from the basal plane of the Si or C terminated faces, in addition to the thermodynamically favored step height. The smallest miscut angle of commercially available SiC wafers is at most 0.1$^\circ$, but it varies throughout the wafer and leads to variations in the step density and orientation. Preparation of well-defined SiC surfaces with regularly spaced steps, by hydrogen etching, can be preserved in a range of growth parameters \cite{oliveira} but generally leads to further step bunching at the graphene growth temperature (T\textgreater$1300^\circ$C in the confinement controlled sublimation method \cite{deHeer29092011} or higher under atmospheric pressure argon \cite{SeyllerNatureMat}). These methods, however, cannot provide a precise alignment of the steps. Another way is to get rid of the steps altogether by step flow at the edges of etched mesas\cite{deHeer29092011}. 

On unpatterned EG on SiC, the steps tend to bunch into unit cell high steps\cite{zangwill}, which is 1 nm high for 4H-SiC. Their spacing is determined by the local miscut angle, which for on-axis wafers is typically in the range 0.1$^{\circ}$ to 0.5$^{\circ}$, giving step spacing of 600nm to 110nm. Under the proper H$_{2}$ etching and growth conditions higher step heights give correspondingly wider terraces, often around 1 $\mu$m wide. \cite{oliveira} However, a naturally formed surface cannot have an arbitrarirly determined step width, and in addition an arbitrarily determined step location and orientation. This means that given step spacings of this size, steps are likely to interfere with a graphene device.

To successfully tailor growth of EG on SiC requires placement of structures that are stable at EG growth temperatures on the SiC surface. Mesas \cite{deHeer29092011} and trenches \cite{Hu} etched into SiC are one solution.  Generally, any surface configuration that would prevent step diffusion on SiC will successfully structure step bunching, including on other materials such as silicon\cite{tanaka,Lee200032}. Deeply etched grid features in these silicon samples enclose the step-flow \cite{tanaka,Lee200032}.

Recognizing that graphene layers hinder step-flow\cite{deHeer29092011}, and that graphitic materials are also high temperature stable, we propose here to enclose the naturally occuring surface steps on SiC with evaporated amorphous carbon (aC). Compared with other refractory materials, amorphous carbon is completely compatible with graphene growth as it will not introduce additional chemical elements. Annealed aC forms nanocrystalline graphite \cite{PhysRevB.67.214202} and is at least as structurally stable as graphene, such that it retains its patterned shape. 

Creating a two-dimensional corral-like enclosure confines the diffusion of the atoms on the surface, such that any preferred step-flow direction will organize the steps into fewer larger steps bunched at one end of the enclosure. See Figure~\ref{schematic} for a schematic cutaway of the step bunching controlled growth. We choose a grid with dimensions that is comparable with the expected distance over which steps will bunch, which is typically a few microns. 

This step-flow control method harnesses the surface free energy minimization such that step bunching is favored at one end of the enclosure. Mobile steps flow by surface diffusion or sublimation of the surface atoms. Step-flow on Si is mediated by evaporation of the Si atoms from the surface.\cite{tanaka,Lee200032}  However, with EG on SiC, the step-flow is a separate mechanism from the graphene growth.\cite{fanming} If graphene is growing, additional step flow due to Si sublimation is possible. Note that unlike in silicon, the step flow on SiC before EG growth conserves mass. Steps accumulate at one end of an individual grid cell according the local miscut angle on the sample surface.

\begin{figure}
\includegraphics[height=4.2in]{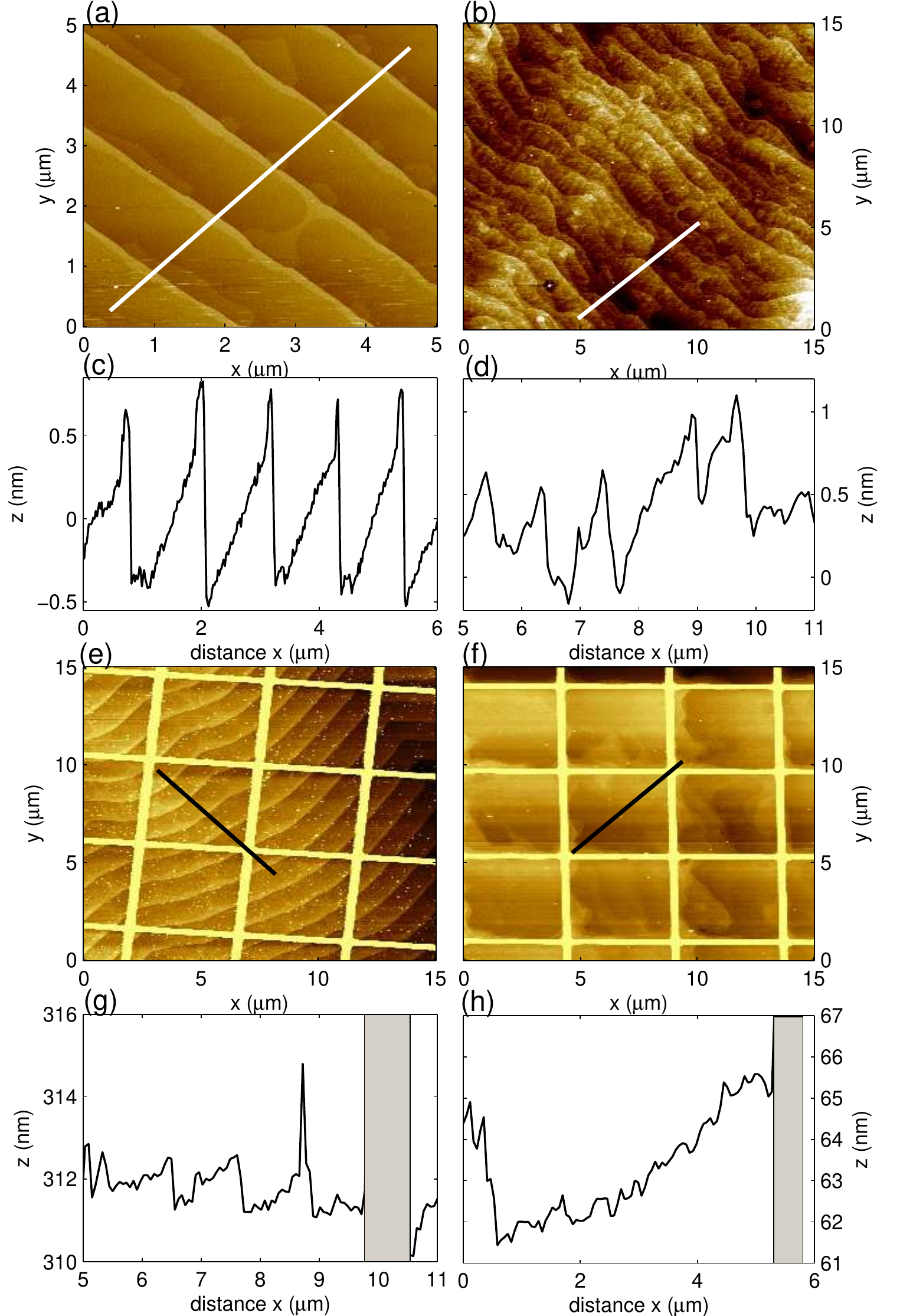}
\caption{AFM images of the surface of hydrogen etched SiC(0001) (a) before and (b) after graphene growth. The corresponding step profiles taken along the step flow (white line in figures (a) and (b)) (c) before and (d) after graphene growth. The AFM images (e) and (f) and corresponding profiles along step flow (g) and (h), respectively,  depict the morphology before and after graphene growth using the amorphous carbon grid.}
\label{graphene topo}
\end{figure}

We prepared EG samples using research grade 4H SiC (0001) and ($000\overline{1})$ substrates cut to 3.5$\times$4.5 mm$^2$ dimensions. These chips came from a nominally on-axis wafer from Cree corporation having a nominal miscut of 0$^\circ$, with an actual miscut angle of up to  0.1$^\circ$. The samples were ultrasonically cleaned for thirty minutes each in acetone and isopropyl alcohol. Three samples were studied: G1, which has monolayer graphene grown on the hydrogen etched Si-face, G2 which is multilayer graphene grown on the polished C-face (see supplemental for images of G2), and B1 which is buffer layer grown on chemical and mechanically polished Si-face. Sample G1 was hydrogen etched in a 40cc/s  ($3\%$ H$_{2}$)/($97\%$ Ar) flowing forming gas mixture at 1450$^\circ$ C at one atm. The hydrogen etching furnace is an inductively heated furnace with a tantalum sample holder. The hydrogen etched surface has well defined SiC steps before the step controlled graphene growth. As seen from Figure ~\ref{graphene topo}, the initial step height on the hydrogen etched sample is about ~1nm, corresponding to 4 SiC unit cells.

Amorphous carbon (aC) is deposited using a Cressington 108A carbon deposition system modified for high vacuum use. Deposition occurs by resistively heating a graphite rod point contact junction to a temperature exceeding the temperature of graphite sublimation. A 3 $\Omega$ junction evaporates with a current 120 A. The aC was uniformly deposited at $10^{-6}$ mbar over all the sample surface.  An in-situ quartz crystal monitors the deposited aC thickness, until a typical a thickness of 20nm is deposited. 

The aC is then patterned into a 5$\times$5 $\mu$m$^2$ square grid pattern using standard electron beam lithography. For this, MaN 2403 negative e-beam resist is spin coated onto the samples at 5000 RPM and baked for 60 seconds at 90$^\circ$ C on a hot plate. E-beam exposure occurs with a cumulative dose of 200 $\mu$C/cm$^2$. The resist is developed for 5 minutes in MF-319 (3$\%$ TMAH) developer. Oxygen plasma RIE etches the unmasked aC. The oxygen is flowed at 4 sccm and RF power of 16 W is applied to the reaction chamber. A typical etch time is 75 seconds. Once the aC is etched through, any overetch into the SiC will introduce a silicon oxide passivation layer to the SiC, which is maximally a few nanometers thick. The oxide layer thickness is self-limiting and is removed during an anneal step as part of the Confinement Controlled Sublimation (CCS) graphene growth process in the furnace \cite{ deHeer29092011}.

Prior to growth, the samples are cleaned overnight at room temperature in an acetone solution to dissolve the remaining resist, and then the samples are rinsed with isopropol alcohol. Samples are grown using Confinement Controlled Sublimation (CCS) in a graphite enclosure in an induction heated furnace. \cite{deHeer29092011} The sample is annealed at 1150$^\circ$ C for 20 minutes to remove silicon dioxide, and then the graphene growth step proceeds.The  Si-face graphene sample G1 was grown at 1600$^\circ$ C for 40 minutes and the buffer layer sample B1 was grown at 1350$^\circ$ C for 30 minutes.

\begin{figure}
\includegraphics[height=4.2in]{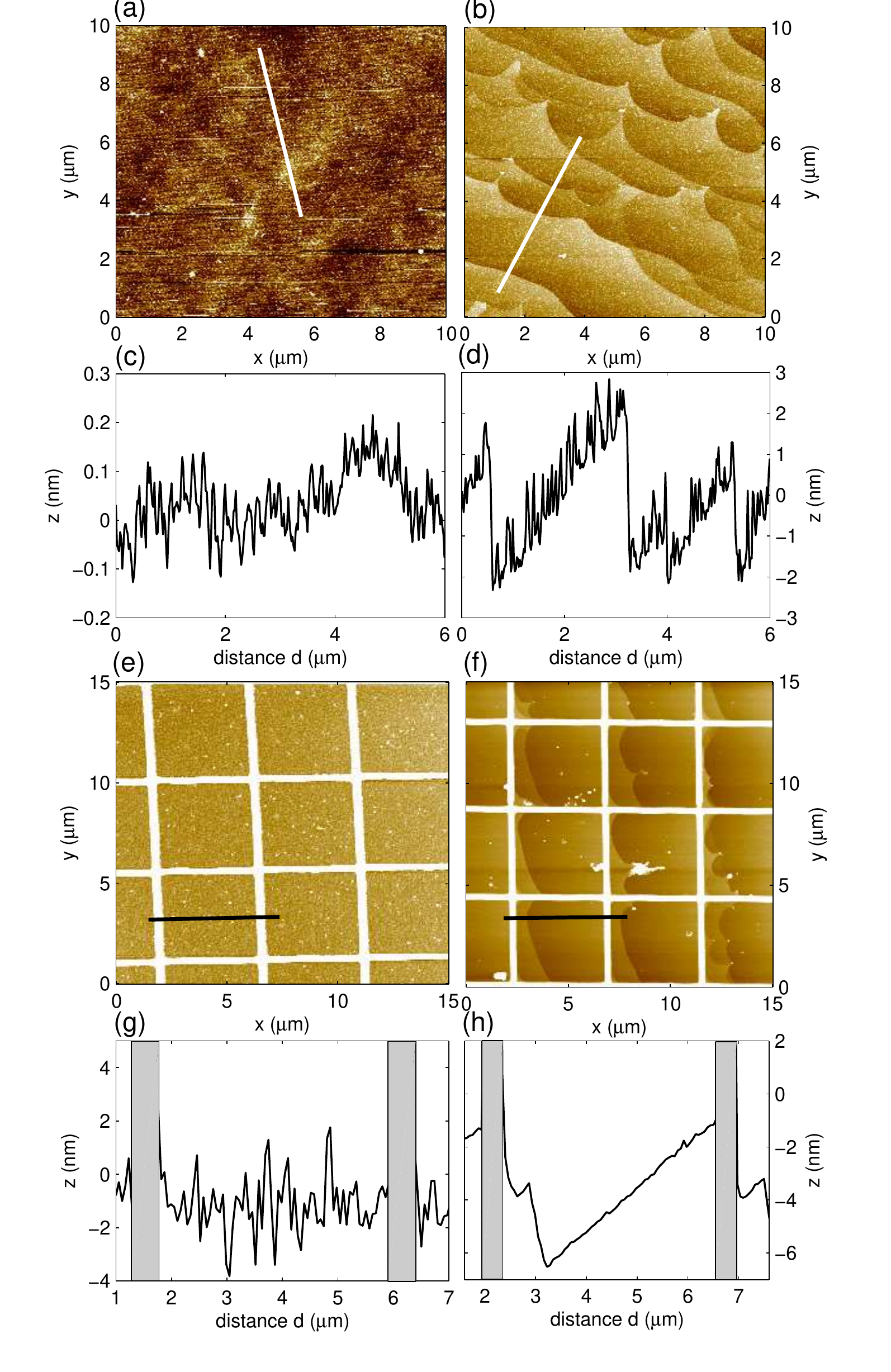}
\caption{As in Fig. (2) but here for sample B1. The SiC surface was chemically and mechanically etched prior to growth. The process of buffer formation without aC grid (a) before and (b) after buffer growth, corresponding profiles (c) and (d). The surface morphology using aC grid (g) before and (h) after buffer growth. Corresponding step profiles along the step flow are shown in figures (g) before and (h) after buffer growth. }
\label{buffer topo}
\end{figure}

Characterization of the surface topography at various stages of process was done using non-contact mode atomic force microscopy (AFM). Before the growth on the unetched surface, there is a nanometer scale roughness in the region within the aC enclosure.(Figure~\ref{graphene topo} and~\ref{buffer topo} part (e))This roughness does not remain after the graphene growth due to removal of the surface oxide layer.  On Figure~\ref{graphene topo} part (e) and (f) note that the orientation of the steps is different in each sub figure. This is due to differing local wafer miscut throughout the sample surface.

  See Figures~\ref{graphene topo} and~\ref{buffer topo} part (f) for the AFM topography after the sample growth. All samples indicate the step morphology aligns to the aC grid.  Larger, nanometer sized aC particles from the aC deposition pin the steps seen in sample B1 away from the gridded area into a pinched shape(Figure ~\ref{buffer topo} (b)). Nanometer sized aC particles are deposited in greater numbers as the carbon rod source is heated to higher temperatures.

Averaging the step widths from the AFM images indicates that the average step width before growth on sample G1 is 0.96 $\mu$m. (Figure ~\ref{graphene topo}, subfigures (a) and (c))  Away from the grid the average step width is 0.22 $\mu$m after growth, and within the grid the step width is increased to 1.9 $\mu$m. On sample B1 the average step width before growth is 0.2 $\mu$m. After growth the average step width away from the grid is 1.4 $\mu$m and within the grid the step width is 3.3 $\mu$m. In both hydrogen etched and non-hydrogen etched cases the presence of the step flow barrier increases the average step width.

Using Non-negative Matrix Factorization (NMF) spectral decomposition to separate the Raman spectra into its principle components\cite{kunc}, the Raman spectra of the graphene on sample G1 may be separated from the aC (see Figure 4). The principle component corresponding to the graphene indicates monolayer growth according to the intensity of the peaks. The principle component that corresponds to the aC indicates nanocrystalline graphite, where the ratio of the D to G peak intensity $\frac{I_D}{I_G}$ = 0.86. This ratio corresponds to a graphitic crystallite size of 20 nm.\cite{cancado:163106} Nanocrystals form upon annealing the aC at graphene growth temperatures. Buffer layer Raman spectra, with a much weaker signal than graphene,  is seen on sample B1. \cite{Fromm,tiberj}

\begin{figure}
\includegraphics[height=4in]{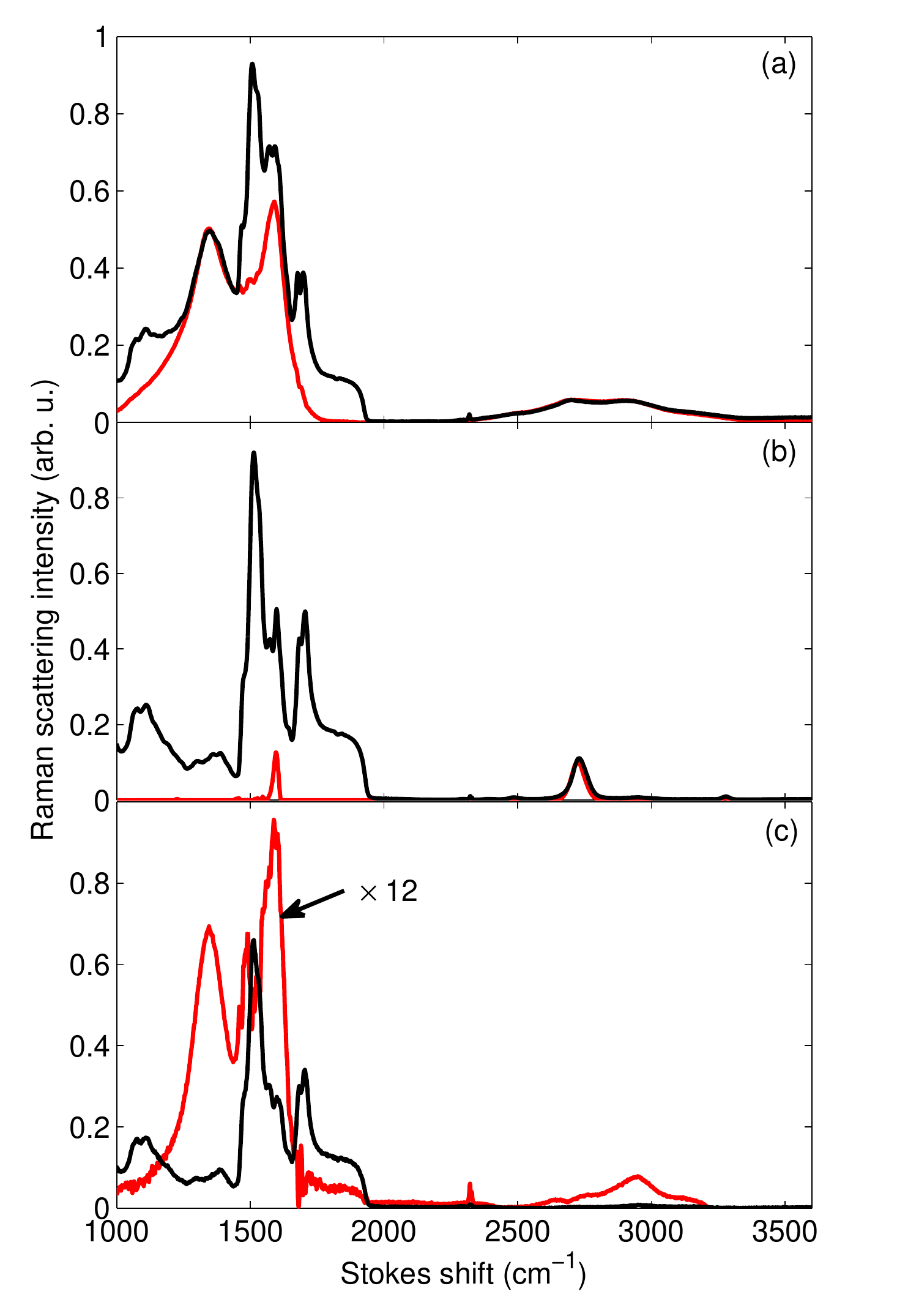}
\caption{Raman scattering spectra (using a $\lambda = $ 532nm laser) of (a) in red: aC and in black: aC and SiC, (b) in red: Sample G1 single layer graphene and in black: graphene and SiC (c) in red: sample B1 buffer layer on a Si-face 4H-SiC, and in black: SiC and buffer.}
\label{Raman}
\end{figure}

Understanding the role that the aC grid structure, growth temperature and SiC wafer miscut play is important for optimal step flow control. For instance, the step flow of the buffer sample B1 appears to be better organized than on the graphene sample G1. The reasons for this difference will require further experiments. We can nevertheless conclude firstly that the step flow indeed involves mass transfer within the aC enclosure only. This is because the total mass of SiC has been conserved before and after graphitization as calculated (The volume of silicon carbide is calculated from step heights and positions before and after growth as well as the carbon contribution to graphene.) from the position of the steps in the AFM images of samples G1 and B1. This mass conservation includes the loss of silicon due to graphene growth (roughly 3 SiC bilayer are required to form one graphene layer). Secondly, the 5$\times$5 $\mu$m$^2$ square grid also appears to be smaller than the distance over which steps will rearrange, particularly on sample B1, since one dominate bunched step is present after growth.

Further step-flow experiments suggested include modifying the aC grid, the wafer miscut, and the growth conditions. Different sized and oriented aC step flow enclosures are in principle possible in order to tailor the step orientation according to the boundary conditions set by the grid.  Additionally, a suitable insulating alternative to aC, like oxides or diamond-like carbon, may make the step flow control even more desirable from a device perspective, since aC is conductive. Performing step flow control on vicinal cut wafers may yield more consistent step bunching, since the total miscut angle will not be as sensitive to the miscut angle uncertainty, unlike on-axis wafers. The step bunching is likely to be greater as well on vicinal cut wafers, thereby enhancing the step flow control.\cite{zangwill} Finally, by increasing the background pressure with an inert gas in the CCS furnace it is possible to raise the onset temperature of graphene growth\cite{deHeer29092011}. This would increase the step flow rates to possibly reach lower surface free energy before graphene forms and therefore inhibits step-flow during graphitization. Lower surface free energy may also be attained by hydrogen etching the sample prior to patterning the grid \cite{oliveira}. It is possible to tune the growth and hydrogen etch procedures to have even larger or specifically tailored step-free areas, in addition to learning more about SiC surface kinetics and graphene growth.

In conclusion, we have shown that evaporated amorphous carbon is effective for structured EG growth and is completely compatible with the CCS growth method. We have demonstrated that SiC step bunching is pinned by an aC cap that acts as a step flow barrier. The step morphology is aligned to the aC grid. Single layer graphene is thereby grown on up to 4 $\mu$m wide step-free areas which locations are predefined by lithography. Further work ought to yield larger step free areas and the possibility of controlling the growth of sidewall graphene on the bunched steps. Careful measurements of step flow control ought to yield deeper information about the graphene growth process on SiC, including mass transport and graphene nucleation rates. Integration of graphene into nanoelectronic processes demands structured growth, such that location and morphology of the graphene are controlled.

We acknowledge financial support from NSF-MRSEC (\#DMR-0820382), AFSOR, the W. M. Keck foundation, and the Partner University Fund. We also thank Andrew Zangwill for helpful advice.


\begin{thebibliography}{28}%
\makeatletter
\providecommand \@ifxundefined [1]{%
 \@ifx{#1\undefined}
}%
\providecommand \@ifnum [1]{%
 \ifnum #1\expandafter \@firstoftwo
 \else \expandafter \@secondoftwo
 \fi
}%
\providecommand \@ifx [1]{%
 \ifx #1\expandafter \@firstoftwo
 \else \expandafter \@secondoftwo
 \fi
}%
\providecommand \natexlab [1]{#1}%
\providecommand \enquote  [1]{``#1''}%
\providecommand \bibnamefont  [1]{#1}%
\providecommand \bibfnamefont [1]{#1}%
\providecommand \citenamefont [1]{#1}%
\providecommand \href@noop [0]{\@secondoftwo}%
\providecommand \href [0]{\begingroup \@sanitize@url \@href}%
\providecommand \@href[1]{\@@startlink{#1}\@@href}%
\providecommand \@@href[1]{\endgroup#1\@@endlink}%
\providecommand \@sanitize@url [0]{\catcode `\\12\catcode `\$12\catcode
  `\&12\catcode `\#12\catcode `\^12\catcode `\_12\catcode `\%12\relax}%
\providecommand \@@startlink[1]{}%
\providecommand \@@endlink[0]{}%
\providecommand \url  [0]{\begingroup\@sanitize@url \@url }%
\providecommand \@url [1]{\endgroup\@href {#1}{\urlprefix }}%
\providecommand \urlprefix  [0]{URL }%
\providecommand \Eprint [0]{\href }%
\providecommand \doibase [0]{http://dx.doi.org/}%
\providecommand \selectlanguage [0]{\@gobble}%
\providecommand \bibinfo  [0]{\@secondoftwo}%
\providecommand \bibfield  [0]{\@secondoftwo}%
\providecommand \translation [1]{[#1]}%
\providecommand \BibitemOpen [0]{}%
\providecommand \bibitemStop [0]{}%
\providecommand \bibitemNoStop [0]{.\EOS\space}%
\providecommand \EOS [0]{\spacefactor3000\relax}%
\providecommand \BibitemShut  [1]{\csname bibitem#1\endcsname}%
\let\auto@bib@innerbib\@empty
\bibitem [{\citenamefont {Berger}\ \emph {et~al.}(2004)\citenamefont {Berger},
  \citenamefont {Song}, \citenamefont {Li}, \citenamefont {Li}, \citenamefont
  {Ogbazghi}, \citenamefont {Feng}, \citenamefont {Dai}, \citenamefont
  {Marchenkov}, \citenamefont {Conrad}, \citenamefont {First},\ and\
  \citenamefont {de~Heer}}]{jphysb2004}%
  \BibitemOpen
  \bibfield  {author} {\bibinfo {author} {\bibfnamefont {C.}~\bibnamefont
  {Berger}}, \bibinfo {author} {\bibfnamefont {Z.}~\bibnamefont {Song}},
  \bibinfo {author} {\bibfnamefont {T.}~\bibnamefont {Li}}, \bibinfo {author}
  {\bibfnamefont {X.}~\bibnamefont {Li}}, \bibinfo {author} {\bibfnamefont
  {A.~Y.}\ \bibnamefont {Ogbazghi}}, \bibinfo {author} {\bibfnamefont
  {R.}~\bibnamefont {Feng}}, \bibinfo {author} {\bibfnamefont {Z.}~\bibnamefont
  {Dai}}, \bibinfo {author} {\bibfnamefont {A.~N.}\ \bibnamefont {Marchenkov}},
  \bibinfo {author} {\bibfnamefont {E.~H.}\ \bibnamefont {Conrad}}, \bibinfo
  {author} {\bibfnamefont {P.~N.}\ \bibnamefont {First}}, \ and\ \bibinfo
  {author} {\bibfnamefont {W.~A.}\ \bibnamefont {de~Heer}},\ }\href {\doibase
  10.1021/jp040650f} {\bibfield  {journal} {\bibinfo  {journal} {The Journal of
  Physical Chemistry B}\ }\textbf {\bibinfo {volume} {108}},\ \bibinfo {pages}
  {19912} (\bibinfo {year} {2004})},\ \Eprint
  {http://arxiv.org/abs/http://pubs.acs.org/doi/pdf/10.1021/jp040650f}
  {http://pubs.acs.org/doi/pdf/10.1021/jp040650f} \BibitemShut {NoStop}%
\bibitem [{\citenamefont {Guo}\ \emph {et~al.}(2013)\citenamefont {Guo},
  \citenamefont {Dong}, \citenamefont {Chakraborty}, \citenamefont {Lourenco},
  \citenamefont {Palmer}, \citenamefont {Hu}, \citenamefont {Ruan},
  \citenamefont {Hankinson}, \citenamefont {Kunc}, \citenamefont {Cressler},
  \citenamefont {Berger},\ and\ \citenamefont {de~Heer}}]{guo2013}%
  \BibitemOpen
  \bibfield  {author} {\bibinfo {author} {\bibfnamefont {Z.}~\bibnamefont
  {Guo}}, \bibinfo {author} {\bibfnamefont {R.}~\bibnamefont {Dong}}, \bibinfo
  {author} {\bibfnamefont {P.~S.}\ \bibnamefont {Chakraborty}}, \bibinfo
  {author} {\bibfnamefont {N.}~\bibnamefont {Lourenco}}, \bibinfo {author}
  {\bibfnamefont {J.}~\bibnamefont {Palmer}}, \bibinfo {author} {\bibfnamefont
  {Y.}~\bibnamefont {Hu}}, \bibinfo {author} {\bibfnamefont {M.}~\bibnamefont
  {Ruan}}, \bibinfo {author} {\bibfnamefont {J.}~\bibnamefont {Hankinson}},
  \bibinfo {author} {\bibfnamefont {J.}~\bibnamefont {Kunc}}, \bibinfo {author}
  {\bibfnamefont {J.~D.}\ \bibnamefont {Cressler}}, \bibinfo {author}
  {\bibfnamefont {C.}~\bibnamefont {Berger}}, \ and\ \bibinfo {author}
  {\bibfnamefont {W.~A.}\ \bibnamefont {de~Heer}},\ }\href {\doibase
  10.1021/nl303587r} {\bibfield  {journal} {\bibinfo  {journal} {Nano Letters}\
  }\textbf {\bibinfo {volume} {13}},\ \bibinfo {pages} {942} (\bibinfo {year}
  {2013})},\ \Eprint
  {http://arxiv.org/abs/http://pubs.acs.org/doi/pdf/10.1021/nl303587r}
  {http://pubs.acs.org/doi/pdf/10.1021/nl303587r} \BibitemShut {NoStop}%
\bibitem [{\citenamefont {Dlubak}\ \emph {et~al.}(2012)\citenamefont {Dlubak},
  \citenamefont {Martin}, \citenamefont {Deranlot}, \citenamefont {Servet},
  \citenamefont {Xavier}, \citenamefont {Mattana}, \citenamefont {Sprinkle},
  \citenamefont {Berger}, \citenamefont {De~Heer}, \citenamefont {Petroff},
  \citenamefont {Anane}, \citenamefont {Seneor},\ and\ \citenamefont
  {Fert}}]{dlubak_highly_2012}%
  \BibitemOpen
  \bibfield  {author} {\bibinfo {author} {\bibfnamefont {B.}~\bibnamefont
  {Dlubak}}, \bibinfo {author} {\bibfnamefont {M.-B.}\ \bibnamefont {Martin}},
  \bibinfo {author} {\bibfnamefont {C.}~\bibnamefont {Deranlot}}, \bibinfo
  {author} {\bibfnamefont {B.}~\bibnamefont {Servet}}, \bibinfo {author}
  {\bibfnamefont {S.}~\bibnamefont {Xavier}}, \bibinfo {author} {\bibfnamefont
  {R.}~\bibnamefont {Mattana}}, \bibinfo {author} {\bibfnamefont
  {M.}~\bibnamefont {Sprinkle}}, \bibinfo {author} {\bibfnamefont
  {C.}~\bibnamefont {Berger}}, \bibinfo {author} {\bibfnamefont {W.~A.}\
  \bibnamefont {De~Heer}}, \bibinfo {author} {\bibfnamefont {F.}~\bibnamefont
  {Petroff}}, \bibinfo {author} {\bibfnamefont {A.}~\bibnamefont {Anane}},
  \bibinfo {author} {\bibfnamefont {P.}~\bibnamefont {Seneor}}, \ and\ \bibinfo
  {author} {\bibfnamefont {A.}~\bibnamefont {Fert}},\ }\href {\doibase
  10.1038/nphys2331} {\bibfield  {journal} {\bibinfo  {journal} {Nat Phys}\
  }\textbf {\bibinfo {volume} {8}},\ \bibinfo {pages} {557} (\bibinfo {year}
  {2012})}\BibitemShut {NoStop}%
\bibitem [{\citenamefont {Baringhaus}\ \emph {et~al.}(2014)\citenamefont
  {Baringhaus}, \citenamefont {Ruan}, \citenamefont {Edler}, \citenamefont
  {Tejeda}, \citenamefont {Sicot}, \citenamefont {{{AminaTaleb-Ibrahimi}}},
  \citenamefont {Li}, \citenamefont {Jiang}, \citenamefont {Conrad},
  \citenamefont {Berger}, \citenamefont {Tegenkamp},\ and\ \citenamefont
  {de~Heer}}]{ballistic}%
  \BibitemOpen
  \bibfield  {author} {\bibinfo {author} {\bibfnamefont {J.}~\bibnamefont
  {Baringhaus}}, \bibinfo {author} {\bibfnamefont {M.}~\bibnamefont {Ruan}},
  \bibinfo {author} {\bibfnamefont {F.}~\bibnamefont {Edler}}, \bibinfo
  {author} {\bibfnamefont {A.}~\bibnamefont {Tejeda}}, \bibinfo {author}
  {\bibfnamefont {M.}~\bibnamefont {Sicot}}, \bibinfo {author} {\bibnamefont
  {{{AminaTaleb-Ibrahimi}}}}, \bibinfo {author} {\bibfnamefont {A.-P.}\
  \bibnamefont {Li}}, \bibinfo {author} {\bibfnamefont {Z.}~\bibnamefont
  {Jiang}}, \bibinfo {author} {\bibfnamefont {E.~H.}\ \bibnamefont {Conrad}},
  \bibinfo {author} {\bibfnamefont {C.}~\bibnamefont {Berger}}, \bibinfo
  {author} {\bibfnamefont {C.}~\bibnamefont {Tegenkamp}}, \ and\ \bibinfo
  {author} {\bibfnamefont {W.~A.}\ \bibnamefont {de~Heer}},\ }\href
  {http://dx.doi.org/10.1038/nature12952} {\bibfield  {journal} {\bibinfo
  {journal} {Nature}\ }\textbf {\bibinfo {volume} {advance online publication}}
  (\bibinfo {year} {2014})}\BibitemShut {NoStop}%
\bibitem [{\citenamefont {{{Sprinkle M.}}}\ \emph {et~al.}(2010)\citenamefont
  {{{Sprinkle M.}}}, \citenamefont {{{Ruan M.}}}, \citenamefont {{{Hu Y.}}},
  \citenamefont {{{Hankinson J.}}}, \citenamefont {{Rubio-{Roy M.}}},
  \citenamefont {{{Zhang B.}}}, \citenamefont {{{Wu X.}}}, \citenamefont
  {{{Berger C.}}},\ and\ \citenamefont {{de {Heer W.} A.}}}]{sprinkle}%
  \BibitemOpen
  \bibfield  {author} {\bibinfo {author} {\bibnamefont {{{Sprinkle M.}}}},
  \bibinfo {author} {\bibnamefont {{{Ruan M.}}}}, \bibinfo {author}
  {\bibnamefont {{{Hu Y.}}}}, \bibinfo {author} {\bibnamefont {{{Hankinson
  J.}}}}, \bibinfo {author} {\bibnamefont {{Rubio-{Roy M.}}}}, \bibinfo
  {author} {\bibnamefont {{{Zhang B.}}}}, \bibinfo {author} {\bibnamefont {{{Wu
  X.}}}}, \bibinfo {author} {\bibnamefont {{{Berger C.}}}}, \ and\ \bibinfo
  {author} {\bibnamefont {{de {Heer W.} A.}}},\ }\href {\doibase
  10.1038/nnano.2010.192} {\bibfield  {journal} {\bibinfo  {journal} {Nat
  Nano}\ }\textbf {\bibinfo {volume} {5}},\ \bibinfo {pages} {727} (\bibinfo
  {year} {2010})}\BibitemShut {NoStop}%
\bibitem [{\citenamefont {Fukidome}\ \emph {et~al.}(2012)\citenamefont
  {Fukidome}, \citenamefont {Kawai}, \citenamefont {Fromm}, \citenamefont
  {Kotsugi}, \citenamefont {Handa}, \citenamefont {Ide}, \citenamefont
  {Ohkouchi}, \citenamefont {Miyashita}, \citenamefont {Enta}, \citenamefont
  {Kinoshita}, \citenamefont {Seyller},\ and\ \citenamefont
  {Suemitsu}}]{handa}%
  \BibitemOpen
  \bibfield  {author} {\bibinfo {author} {\bibfnamefont {H.}~\bibnamefont
  {Fukidome}}, \bibinfo {author} {\bibfnamefont {Y.}~\bibnamefont {Kawai}},
  \bibinfo {author} {\bibfnamefont {F.}~\bibnamefont {Fromm}}, \bibinfo
  {author} {\bibfnamefont {M.}~\bibnamefont {Kotsugi}}, \bibinfo {author}
  {\bibfnamefont {H.}~\bibnamefont {Handa}}, \bibinfo {author} {\bibfnamefont
  {T.}~\bibnamefont {Ide}}, \bibinfo {author} {\bibfnamefont {T.}~\bibnamefont
  {Ohkouchi}}, \bibinfo {author} {\bibfnamefont {H.}~\bibnamefont {Miyashita}},
  \bibinfo {author} {\bibfnamefont {Y.}~\bibnamefont {Enta}}, \bibinfo {author}
  {\bibfnamefont {T.}~\bibnamefont {Kinoshita}}, \bibinfo {author}
  {\bibfnamefont {T.}~\bibnamefont {Seyller}}, \ and\ \bibinfo {author}
  {\bibfnamefont {M.}~\bibnamefont {Suemitsu}},\ }\href {\doibase
  http://dx.doi.org/10.1063/1.4740271} {\bibfield  {journal} {\bibinfo
  {journal} {Applied Physics Letters}\ }\textbf {\bibinfo {volume} {101}},\
  \bibinfo {eid} {041605} (\bibinfo {year} {2012})}\BibitemShut {NoStop}%
\bibitem [{\citenamefont {Rubio-Roy}\ \emph {et~al.}(2010)\citenamefont
  {Rubio-Roy}, \citenamefont {Zaman}, \citenamefont {Hu}, \citenamefont
  {Berger}, \citenamefont {Moseley}, \citenamefont {Meindl},\ and\
  \citenamefont {de~Heer}}]{roy10}%
  \BibitemOpen
  \bibfield  {author} {\bibinfo {author} {\bibfnamefont {M.}~\bibnamefont
  {Rubio-Roy}}, \bibinfo {author} {\bibfnamefont {F.}~\bibnamefont {Zaman}},
  \bibinfo {author} {\bibfnamefont {Y.}~\bibnamefont {Hu}}, \bibinfo {author}
  {\bibfnamefont {C.}~\bibnamefont {Berger}}, \bibinfo {author} {\bibfnamefont
  {M.~W.}\ \bibnamefont {Moseley}}, \bibinfo {author} {\bibfnamefont {J.~D.}\
  \bibnamefont {Meindl}}, \ and\ \bibinfo {author} {\bibfnamefont {W.~A.}\
  \bibnamefont {de~Heer}},\ }\href {\doibase
  http://dx.doi.org/10.1063/1.3334683} {\bibfield  {journal} {\bibinfo
  {journal} {Applied Physics Letters}\ }\textbf {\bibinfo {volume} {96}},\
  \bibinfo {eid} {082112} (\bibinfo {year} {2010})}\BibitemShut {NoStop}%
\bibitem [{\citenamefont {{Puybaret}}\ \emph {et~al.}(2013)\citenamefont
  {{Puybaret}}, \citenamefont {{Hankinson}}, \citenamefont {{Ougazzaden}},
  \citenamefont {{Voss}}, \citenamefont {{Berger}},\ and\ \citenamefont {{de
  Heer}}}]{puybaret}%
  \BibitemOpen
  \bibfield  {author} {\bibinfo {author} {\bibfnamefont {R.}~\bibnamefont
  {{Puybaret}}}, \bibinfo {author} {\bibfnamefont {J.}~\bibnamefont
  {{Hankinson}}}, \bibinfo {author} {\bibfnamefont {A.}~\bibnamefont
  {{Ougazzaden}}}, \bibinfo {author} {\bibfnamefont {P.~L.}\ \bibnamefont
  {{Voss}}}, \bibinfo {author} {\bibfnamefont {C.}~\bibnamefont {{Berger}}}, \
  and\ \bibinfo {author} {\bibfnamefont {W.~A.}\ \bibnamefont {{de Heer}}},\
  }\href@noop {} {\bibfield  {journal} {\bibinfo  {journal} {ArXiv e-prints}\ }
  (\bibinfo {year} {2013})},\ \Eprint {http://arxiv.org/abs/1307.6197}
  {arXiv:1307.6197 [cond-mat.mtrl-sci]} \BibitemShut {NoStop}%
\bibitem [{\citenamefont {Camara}\ \emph {et~al.}(2009)\citenamefont {Camara},
  \citenamefont {Huntzinger}, \citenamefont {Rius}, \citenamefont {Tiberj},
  \citenamefont {Mestres}, \citenamefont {P\'erez-Murano}, \citenamefont
  {Godignon},\ and\ \citenamefont {Camassel}}]{camara}%
  \BibitemOpen
  \bibfield  {author} {\bibinfo {author} {\bibfnamefont {N.}~\bibnamefont
  {Camara}}, \bibinfo {author} {\bibfnamefont {J.-R.}\ \bibnamefont
  {Huntzinger}}, \bibinfo {author} {\bibfnamefont {G.}~\bibnamefont {Rius}},
  \bibinfo {author} {\bibfnamefont {A.}~\bibnamefont {Tiberj}}, \bibinfo
  {author} {\bibfnamefont {N.}~\bibnamefont {Mestres}}, \bibinfo {author}
  {\bibfnamefont {F.}~\bibnamefont {P\'erez-Murano}}, \bibinfo {author}
  {\bibfnamefont {P.}~\bibnamefont {Godignon}}, \ and\ \bibinfo {author}
  {\bibfnamefont {J.}~\bibnamefont {Camassel}},\ }\href {\doibase
  10.1103/PhysRevB.80.125410} {\bibfield  {journal} {\bibinfo  {journal} {Phys.
  Rev. B}\ }\textbf {\bibinfo {volume} {80}},\ \bibinfo {pages} {125410}
  (\bibinfo {year} {2009})}\BibitemShut {NoStop}%
\bibitem [{\citenamefont {Safron}\ \emph {et~al.}(2012)\citenamefont {Safron},
  \citenamefont {Kim}, \citenamefont {Gopalan},\ and\ \citenamefont
  {Arnold}}]{ADMA:ADMA201104195}%
  \BibitemOpen
  \bibfield  {author} {\bibinfo {author} {\bibfnamefont {N.~S.}\ \bibnamefont
  {Safron}}, \bibinfo {author} {\bibfnamefont {M.}~\bibnamefont {Kim}},
  \bibinfo {author} {\bibfnamefont {P.}~\bibnamefont {Gopalan}}, \ and\
  \bibinfo {author} {\bibfnamefont {M.~S.}\ \bibnamefont {Arnold}},\ }\href
  {\doibase 10.1002/adma.201104195} {\bibfield  {journal} {\bibinfo  {journal}
  {Advanced Materials}\ }\textbf {\bibinfo {volume} {24}},\ \bibinfo {pages}
  {1041} (\bibinfo {year} {2012})}\BibitemShut {NoStop}%
\bibitem [{\citenamefont {Cai}\ \emph {et~al.}(2010)\citenamefont {Cai},
  \citenamefont {Ruffieux}, \citenamefont {Jaafar}, \citenamefont {Bieri},
  \citenamefont {Braun}, \citenamefont {Blankenburg}, \citenamefont {Muoth},
  \citenamefont {Seitsonen}, \citenamefont {Saleh}, \citenamefont {Feng},
  \citenamefont {Mullen},\ and\ \citenamefont {Fasel}}]{caiatomically2010}%
  \BibitemOpen
  \bibfield  {author} {\bibinfo {author} {\bibfnamefont {J.}~\bibnamefont
  {Cai}}, \bibinfo {author} {\bibfnamefont {P.}~\bibnamefont {Ruffieux}},
  \bibinfo {author} {\bibfnamefont {R.}~\bibnamefont {Jaafar}}, \bibinfo
  {author} {\bibfnamefont {M.}~\bibnamefont {Bieri}}, \bibinfo {author}
  {\bibfnamefont {T.}~\bibnamefont {Braun}}, \bibinfo {author} {\bibfnamefont
  {S.}~\bibnamefont {Blankenburg}}, \bibinfo {author} {\bibfnamefont
  {M.}~\bibnamefont {Muoth}}, \bibinfo {author} {\bibfnamefont {A.~P.}\
  \bibnamefont {Seitsonen}}, \bibinfo {author} {\bibfnamefont {M.}~\bibnamefont
  {Saleh}}, \bibinfo {author} {\bibfnamefont {X.}~\bibnamefont {Feng}},
  \bibinfo {author} {\bibfnamefont {K.}~\bibnamefont {Mullen}}, \ and\ \bibinfo
  {author} {\bibfnamefont {R.}~\bibnamefont {Fasel}},\ }\href {\doibase
  10.1038/nature09211} {\bibfield  {journal} {\bibinfo  {journal} {Nature}\
  }\textbf {\bibinfo {volume} {466}},\ \bibinfo {pages} {470} (\bibinfo {year}
  {2010})}\BibitemShut {NoStop}%
\bibitem [{\citenamefont {Dong}\ \emph {et~al.}(2014)\citenamefont {Dong},
  \citenamefont {Guo}, \citenamefont {Palmer}, \citenamefont {Hu},
  \citenamefont {Ruan}, \citenamefont {Hankinson}, \citenamefont {Kunc},
  \citenamefont {Bhattacharya}, \citenamefont {Berger},\ and\ \citenamefont
  {de~Heer}}]{dong}%
  \BibitemOpen
  \bibfield  {author} {\bibinfo {author} {\bibfnamefont {R.}~\bibnamefont
  {Dong}}, \bibinfo {author} {\bibfnamefont {Z.}~\bibnamefont {Guo}}, \bibinfo
  {author} {\bibfnamefont {J.}~\bibnamefont {Palmer}}, \bibinfo {author}
  {\bibfnamefont {Y.}~\bibnamefont {Hu}}, \bibinfo {author} {\bibfnamefont
  {M.}~\bibnamefont {Ruan}}, \bibinfo {author} {\bibfnamefont {J.}~\bibnamefont
  {Hankinson}}, \bibinfo {author} {\bibfnamefont {J.}~\bibnamefont {Kunc}},
  \bibinfo {author} {\bibfnamefont {S.~K.}\ \bibnamefont {Bhattacharya}},
  \bibinfo {author} {\bibfnamefont {C.}~\bibnamefont {Berger}}, \ and\ \bibinfo
  {author} {\bibfnamefont {W.~A.}\ \bibnamefont {de~Heer}},\ }\href
  {http://stacks.iop.org/0022-3727/47/i=9/a=094001} {\bibfield  {journal}
  {\bibinfo  {journal} {Journal of Physics D: Applied Physics}\ }\textbf
  {\bibinfo {volume} {47}},\ \bibinfo {pages} {094001} (\bibinfo {year}
  {2014})}\BibitemShut {NoStop}%
\bibitem [{\citenamefont {Krach}\ \emph {et~al.}(2012)\citenamefont {Krach},
  \citenamefont {Hertel}, \citenamefont {Waldmann}, \citenamefont {Jobst},
  \citenamefont {Krieger}, \citenamefont {Reshanov}, \citenamefont {Schöner},\
  and\ \citenamefont {Weber}}]{weber}%
  \BibitemOpen
  \bibfield  {author} {\bibinfo {author} {\bibfnamefont {F.}~\bibnamefont
  {Krach}}, \bibinfo {author} {\bibfnamefont {S.}~\bibnamefont {Hertel}},
  \bibinfo {author} {\bibfnamefont {D.}~\bibnamefont {Waldmann}}, \bibinfo
  {author} {\bibfnamefont {J.}~\bibnamefont {Jobst}}, \bibinfo {author}
  {\bibfnamefont {M.}~\bibnamefont {Krieger}}, \bibinfo {author} {\bibfnamefont
  {S.}~\bibnamefont {Reshanov}}, \bibinfo {author} {\bibfnamefont
  {A.}~\bibnamefont {Schöner}}, \ and\ \bibinfo {author} {\bibfnamefont
  {H.~B.}\ \bibnamefont {Weber}},\ }\href {\doibase
  http://dx.doi.org/10.1063/1.3695157} {\bibfield  {journal} {\bibinfo
  {journal} {Applied Physics Letters}\ }\textbf {\bibinfo {volume} {100}},\
  \bibinfo {eid} {122102} (\bibinfo {year} {2012})}\BibitemShut {NoStop}%
\bibitem [{\citenamefont {Ohta}\ \emph {et~al.}(2010)\citenamefont {Ohta},
  \citenamefont {Bartelt}, \citenamefont {Nie}, \citenamefont {Th\"urmer},\
  and\ \citenamefont {Kellogg}}]{Ohta}%
  \BibitemOpen
  \bibfield  {author} {\bibinfo {author} {\bibfnamefont {T.}~\bibnamefont
  {Ohta}}, \bibinfo {author} {\bibfnamefont {N.~C.}\ \bibnamefont {Bartelt}},
  \bibinfo {author} {\bibfnamefont {S.}~\bibnamefont {Nie}}, \bibinfo {author}
  {\bibfnamefont {K.}~\bibnamefont {Th\"urmer}}, \ and\ \bibinfo {author}
  {\bibfnamefont {G.~L.}\ \bibnamefont {Kellogg}},\ }\href {\doibase
  10.1103/PhysRevB.81.121411} {\bibfield  {journal} {\bibinfo  {journal} {Phys.
  Rev. B}\ }\textbf {\bibinfo {volume} {81}},\ \bibinfo {pages} {121411}
  (\bibinfo {year} {2010})}\BibitemShut {NoStop}%
\bibitem [{\citenamefont {Ming}\ and\ \citenamefont
  {Zangwill}(2012)}]{fanming}%
  \BibitemOpen
  \bibfield  {author} {\bibinfo {author} {\bibfnamefont {F.}~\bibnamefont
  {Ming}}\ and\ \bibinfo {author} {\bibfnamefont {A.}~\bibnamefont
  {Zangwill}},\ }\href {http://stacks.iop.org/0022-3727/45/i=15/a=154007}
  {\bibfield  {journal} {\bibinfo  {journal} {Journal of Physics D: Applied
  Physics}\ }\textbf {\bibinfo {volume} {45}},\ \bibinfo {pages} {154007}
  (\bibinfo {year} {2012})}\BibitemShut {NoStop}%
\bibitem [{\citenamefont {Ji}\ \emph {et~al.}(2012)\citenamefont {Ji},
  \citenamefont {Hannon}, \citenamefont {Tromp}, \citenamefont {Perebeinos},
  \citenamefont {Tersoff},\ and\ \citenamefont {Ross}}]{ross}%
  \BibitemOpen
  \bibfield  {author} {\bibinfo {author} {\bibfnamefont {S.-H.}\ \bibnamefont
  {Ji}}, \bibinfo {author} {\bibfnamefont {J.~B.}\ \bibnamefont {Hannon}},
  \bibinfo {author} {\bibfnamefont {R.~M.}\ \bibnamefont {Tromp}}, \bibinfo
  {author} {\bibfnamefont {V.}~\bibnamefont {Perebeinos}}, \bibinfo {author}
  {\bibfnamefont {J.}~\bibnamefont {Tersoff}}, \ and\ \bibinfo {author}
  {\bibfnamefont {F.~M.}\ \bibnamefont {Ross}},\ }\href {\doibase
  10.1038/nmat3170} {\bibfield  {journal} {\bibinfo  {journal} {Nat Mater}\
  }\textbf {\bibinfo {volume} {11}},\ \bibinfo {pages} {114} (\bibinfo {year}
  {2012})}\BibitemShut {NoStop}%
\bibitem [{\citenamefont {Oliveira}\ \emph {et~al.}(2011)\citenamefont
  {Oliveira}, \citenamefont {Schumann}, \citenamefont {Ramsteiner},
  \citenamefont {Lopes},\ and\ \citenamefont {Riechert}}]{oliveira}%
  \BibitemOpen
  \bibfield  {author} {\bibinfo {author} {\bibfnamefont {M.~H.}\ \bibnamefont
  {Oliveira}}, \bibinfo {author} {\bibfnamefont {T.}~\bibnamefont {Schumann}},
  \bibinfo {author} {\bibfnamefont {M.}~\bibnamefont {Ramsteiner}}, \bibinfo
  {author} {\bibfnamefont {J.~M.~J.}\ \bibnamefont {Lopes}}, \ and\ \bibinfo
  {author} {\bibfnamefont {H.}~\bibnamefont {Riechert}},\ }\href {\doibase
  http://dx.doi.org/10.1063/1.3638058} {\bibfield  {journal} {\bibinfo
  {journal} {Applied Physics Letters}\ }\textbf {\bibinfo {volume} {99}},\
  \bibinfo {eid} {111901} (\bibinfo {year} {2011})}\BibitemShut {NoStop}%
\bibitem [{\citenamefont {de~Heer}\ \emph {et~al.}(2011)\citenamefont
  {de~Heer}, \citenamefont {Berger}, \citenamefont {Ruan}, \citenamefont
  {Sprinkle}, \citenamefont {Li}, \citenamefont {Hu}, \citenamefont {Zhang},
  \citenamefont {Hankinson},\ and\ \citenamefont {Conrad}}]{deHeer29092011}%
  \BibitemOpen
  \bibfield  {author} {\bibinfo {author} {\bibfnamefont {W.~A.}\ \bibnamefont
  {de~Heer}}, \bibinfo {author} {\bibfnamefont {C.}~\bibnamefont {Berger}},
  \bibinfo {author} {\bibfnamefont {M.}~\bibnamefont {Ruan}}, \bibinfo {author}
  {\bibfnamefont {M.}~\bibnamefont {Sprinkle}}, \bibinfo {author}
  {\bibfnamefont {X.}~\bibnamefont {Li}}, \bibinfo {author} {\bibfnamefont
  {Y.}~\bibnamefont {Hu}}, \bibinfo {author} {\bibfnamefont {B.}~\bibnamefont
  {Zhang}}, \bibinfo {author} {\bibfnamefont {J.}~\bibnamefont {Hankinson}}, \
  and\ \bibinfo {author} {\bibfnamefont {E.}~\bibnamefont {Conrad}},\ }\href
  {\doibase 10.1073/pnas.1105113108} {\bibfield  {journal} {\bibinfo  {journal}
  {Proceedings of the National Academy of Sciences}\ }\textbf {\bibinfo
  {volume} {108}},\ \bibinfo {pages} {16900} (\bibinfo {year} {2011})},\
  \Eprint
  {http://arxiv.org/abs/http://www.pnas.org/content/108/41/16900.full.pdf+html}
  {http://www.pnas.org/content/108/41/16900.full.pdf+html} \BibitemShut
  {NoStop}%
\bibitem [{\citenamefont {Emtsev}\ \emph {et~al.}(2009)\citenamefont {Emtsev},
  \citenamefont {Bostwick}, \citenamefont {Horn}, \citenamefont {Jobst},
  \citenamefont {Kellogg}, \citenamefont {Ley}, \citenamefont {{McChesney}},
  \citenamefont {Ohta}, \citenamefont {Reshanov}, \citenamefont {Rohrl},
  \citenamefont {Rotenberg}, \citenamefont {Schmid}, \citenamefont {Waldmann},
  \citenamefont {Weber},\ and\ \citenamefont {Seyller}}]{SeyllerNatureMat}%
  \BibitemOpen
  \bibfield  {author} {\bibinfo {author} {\bibfnamefont {K.~V.}\ \bibnamefont
  {Emtsev}}, \bibinfo {author} {\bibfnamefont {A.}~\bibnamefont {Bostwick}},
  \bibinfo {author} {\bibfnamefont {K.}~\bibnamefont {Horn}}, \bibinfo {author}
  {\bibfnamefont {J.}~\bibnamefont {Jobst}}, \bibinfo {author} {\bibfnamefont
  {G.~L.}\ \bibnamefont {Kellogg}}, \bibinfo {author} {\bibfnamefont
  {L.}~\bibnamefont {Ley}}, \bibinfo {author} {\bibfnamefont {J.~L.}\
  \bibnamefont {{McChesney}}}, \bibinfo {author} {\bibfnamefont
  {T.}~\bibnamefont {Ohta}}, \bibinfo {author} {\bibfnamefont {S.~A.}\
  \bibnamefont {Reshanov}}, \bibinfo {author} {\bibfnamefont {J.}~\bibnamefont
  {Rohrl}}, \bibinfo {author} {\bibfnamefont {E.}~\bibnamefont {Rotenberg}},
  \bibinfo {author} {\bibfnamefont {A.~K.}\ \bibnamefont {Schmid}}, \bibinfo
  {author} {\bibfnamefont {D.}~\bibnamefont {Waldmann}}, \bibinfo {author}
  {\bibfnamefont {H.~B.}\ \bibnamefont {Weber}}, \ and\ \bibinfo {author}
  {\bibfnamefont {T.}~\bibnamefont {Seyller}},\ }\href {\doibase
  10.1038/nmat2382} {\bibfield  {journal} {\bibinfo  {journal} {Nat Mater}\
  }\textbf {\bibinfo {volume} {8}},\ \bibinfo {pages} {203} (\bibinfo {year}
  {2009})}\BibitemShut {NoStop}%
\bibitem [{\citenamefont {Borovikov}\ and\ \citenamefont
  {Zangwill}(2009)}]{zangwill}%
  \BibitemOpen
  \bibfield  {author} {\bibinfo {author} {\bibfnamefont {V.}~\bibnamefont
  {Borovikov}}\ and\ \bibinfo {author} {\bibfnamefont {A.}~\bibnamefont
  {Zangwill}},\ }\href {\doibase 10.1103/PhysRevB.79.245413} {\bibfield
  {journal} {\bibinfo  {journal} {Phys. Rev. B}\ }\textbf {\bibinfo {volume}
  {79}},\ \bibinfo {pages} {245413} (\bibinfo {year} {2009})}\BibitemShut
  {NoStop}%
\bibitem [{\citenamefont {Hu}\ \emph {et~al.}(2012)\citenamefont {Hu},
  \citenamefont {Ruan}, \citenamefont {Guo}, \citenamefont {Dong},
  \citenamefont {Palmer}, \citenamefont {Hankinson}, \citenamefont {Berger},\
  and\ \citenamefont {de~Heer}}]{Hu}%
  \BibitemOpen
  \bibfield  {author} {\bibinfo {author} {\bibfnamefont {Y.}~\bibnamefont
  {Hu}}, \bibinfo {author} {\bibfnamefont {M.}~\bibnamefont {Ruan}}, \bibinfo
  {author} {\bibfnamefont {Z.}~\bibnamefont {Guo}}, \bibinfo {author}
  {\bibfnamefont {R.}~\bibnamefont {Dong}}, \bibinfo {author} {\bibfnamefont
  {J.}~\bibnamefont {Palmer}}, \bibinfo {author} {\bibfnamefont
  {J.}~\bibnamefont {Hankinson}}, \bibinfo {author} {\bibfnamefont
  {C.}~\bibnamefont {Berger}}, \ and\ \bibinfo {author} {\bibfnamefont {W.~A.}\
  \bibnamefont {de~Heer}},\ }\href
  {http://stacks.iop.org/0022-3727/45/i=15/a=154010} {\bibfield  {journal}
  {\bibinfo  {journal} {Journal of Physics D: Applied Physics}\ }\textbf
  {\bibinfo {volume} {45}},\ \bibinfo {pages} {154010} (\bibinfo {year}
  {2012})}\BibitemShut {NoStop}%
\bibitem [{\citenamefont {Tanaka}\ \emph {et~al.}(1996)\citenamefont {Tanaka},
  \citenamefont {Umbach}, \citenamefont {Blakely}, \citenamefont {Tromp},\ and\
  \citenamefont {Mankos}}]{tanaka}%
  \BibitemOpen
  \bibfield  {author} {\bibinfo {author} {\bibfnamefont {S.}~\bibnamefont
  {Tanaka}}, \bibinfo {author} {\bibfnamefont {C.~C.}\ \bibnamefont {Umbach}},
  \bibinfo {author} {\bibfnamefont {J.~M.}\ \bibnamefont {Blakely}}, \bibinfo
  {author} {\bibfnamefont {R.~M.}\ \bibnamefont {Tromp}}, \ and\ \bibinfo
  {author} {\bibfnamefont {M.}~\bibnamefont {Mankos}},\ }\href@noop {}
  {\bibfield  {journal} {\bibinfo  {journal} {Applied Physics Letters}\
  }\textbf {\bibinfo {volume} {69}} (\bibinfo {year} {1996})}\BibitemShut
  {NoStop}%
\bibitem [{\citenamefont {Lee}\ and\ \citenamefont
  {Blakely}(2000)}]{Lee200032}%
  \BibitemOpen
  \bibfield  {author} {\bibinfo {author} {\bibfnamefont {D.}~\bibnamefont
  {Lee}}\ and\ \bibinfo {author} {\bibfnamefont {J.}~\bibnamefont {Blakely}},\
  }\href {\doibase http://dx.doi.org/10.1016/S0039-6028(99)01034-1} {\bibfield
  {journal} {\bibinfo  {journal} {Surface Science}\ }\textbf {\bibinfo {volume}
  {445}},\ \bibinfo {pages} {32 } (\bibinfo {year} {2000})}\BibitemShut
  {NoStop}%
\bibitem [{\citenamefont {Takai}\ \emph {et~al.}(2003)\citenamefont {Takai},
  \citenamefont {Oga}, \citenamefont {Sato}, \citenamefont {Enoki},
  \citenamefont {Ohki}, \citenamefont {Taomoto}, \citenamefont {Suenaga},\ and\
  \citenamefont {Iijima}}]{PhysRevB.67.214202}%
  \BibitemOpen
  \bibfield  {author} {\bibinfo {author} {\bibfnamefont {K.}~\bibnamefont
  {Takai}}, \bibinfo {author} {\bibfnamefont {M.}~\bibnamefont {Oga}}, \bibinfo
  {author} {\bibfnamefont {H.}~\bibnamefont {Sato}}, \bibinfo {author}
  {\bibfnamefont {T.}~\bibnamefont {Enoki}}, \bibinfo {author} {\bibfnamefont
  {Y.}~\bibnamefont {Ohki}}, \bibinfo {author} {\bibfnamefont {A.}~\bibnamefont
  {Taomoto}}, \bibinfo {author} {\bibfnamefont {K.}~\bibnamefont {Suenaga}}, \
  and\ \bibinfo {author} {\bibfnamefont {S.}~\bibnamefont {Iijima}},\ }\href
  {\doibase 10.1103/PhysRevB.67.214202} {\bibfield  {journal} {\bibinfo
  {journal} {Phys. Rev. B}\ }\textbf {\bibinfo {volume} {67}},\ \bibinfo
  {pages} {214202} (\bibinfo {year} {2003})}\BibitemShut {NoStop}%
\bibitem [{\citenamefont {Kunc}\ \emph {et~al.}(2013)\citenamefont {Kunc},
  \citenamefont {Hu}, \citenamefont {Palmer}, \citenamefont {Berger},\ and\
  \citenamefont {de~Heer}}]{kunc}%
  \BibitemOpen
  \bibfield  {author} {\bibinfo {author} {\bibfnamefont {J.}~\bibnamefont
  {Kunc}}, \bibinfo {author} {\bibfnamefont {Y.}~\bibnamefont {Hu}}, \bibinfo
  {author} {\bibfnamefont {J.}~\bibnamefont {Palmer}}, \bibinfo {author}
  {\bibfnamefont {C.}~\bibnamefont {Berger}}, \ and\ \bibinfo {author}
  {\bibfnamefont {W.~A.}\ \bibnamefont {de~Heer}},\ }\href {\doibase
  http://dx.doi.org/10.1063/1.4830374} {\bibfield  {journal} {\bibinfo
  {journal} {Applied Physics Letters}\ }\textbf {\bibinfo {volume} {103}},\
  \bibinfo {eid} {201911} (\bibinfo {year} {2013})}\BibitemShut {NoStop}%
\bibitem [{\citenamefont {Can\c{c}ado}\ \emph {et~al.}(2006)\citenamefont
  {Can\c{c}ado}, \citenamefont {Takai}, \citenamefont {Enoki}, \citenamefont
  {Endo}, \citenamefont {Kim}, \citenamefont {Mizusaki}, \citenamefont {Jorio},
  \citenamefont {Coelho}, \citenamefont {{a}es Paniago},\ and\ \citenamefont
  {Pimenta}}]{cancado:163106}%
  \BibitemOpen
  \bibfield  {author} {\bibinfo {author} {\bibfnamefont {L.~G.}\ \bibnamefont
  {Can\c{c}ado}}, \bibinfo {author} {\bibfnamefont {K.}~\bibnamefont {Takai}},
  \bibinfo {author} {\bibfnamefont {T.}~\bibnamefont {Enoki}}, \bibinfo
  {author} {\bibfnamefont {M.}~\bibnamefont {Endo}}, \bibinfo {author}
  {\bibfnamefont {Y.~A.}\ \bibnamefont {Kim}}, \bibinfo {author} {\bibfnamefont
  {H.}~\bibnamefont {Mizusaki}}, \bibinfo {author} {\bibfnamefont
  {A.}~\bibnamefont {Jorio}}, \bibinfo {author} {\bibfnamefont {L.~N.}\
  \bibnamefont {Coelho}}, \bibinfo {author} {\bibfnamefont {R.~M.}\
  \bibnamefont {{a}es Paniago}}, \ and\ \bibinfo {author} {\bibfnamefont
  {M.~A.}\ \bibnamefont {Pimenta}},\ }\href {\doibase 10.1063/1.2196057}
  {\bibfield  {journal} {\bibinfo  {journal} {Applied Physics Letters}\
  }\textbf {\bibinfo {volume} {88}},\ \bibinfo {eid} {163106} (\bibinfo {year}
  {2006})}\BibitemShut {NoStop}%
\bibitem [{\citenamefont {Fromm}\ \emph {et~al.}(2013)\citenamefont {Fromm},
  \citenamefont {Jr}, \citenamefont {Molina-Sánchez}, \citenamefont
  {Hundhausen}, \citenamefont {Lopes}, \citenamefont {Riechert}, \citenamefont
  {Wirtz},\ and\ \citenamefont {Seyller}}]{Fromm}%
  \BibitemOpen
  \bibfield  {author} {\bibinfo {author} {\bibfnamefont {F.}~\bibnamefont
  {Fromm}}, \bibinfo {author} {\bibfnamefont {M.~H.~O.}\ \bibnamefont {Jr}},
  \bibinfo {author} {\bibfnamefont {A.}~\bibnamefont {Molina-Sánchez}},
  \bibinfo {author} {\bibfnamefont {M.}~\bibnamefont {Hundhausen}}, \bibinfo
  {author} {\bibfnamefont {J.~M.~J.}\ \bibnamefont {Lopes}}, \bibinfo {author}
  {\bibfnamefont {H.}~\bibnamefont {Riechert}}, \bibinfo {author}
  {\bibfnamefont {L.}~\bibnamefont {Wirtz}}, \ and\ \bibinfo {author}
  {\bibfnamefont {T.}~\bibnamefont {Seyller}},\ }\href
  {http://stacks.iop.org/1367-2630/15/i=4/a=043031} {\bibfield  {journal}
  {\bibinfo  {journal} {New Journal of Physics}\ }\textbf {\bibinfo {volume}
  {15}},\ \bibinfo {pages} {043031} (\bibinfo {year} {2013})}\BibitemShut
  {NoStop}%
\bibitem [{\citenamefont {{Tiberj}}\ \emph {et~al.}(2012)\citenamefont
  {{Tiberj}}, \citenamefont {{Huntzinger}}, \citenamefont {{Camara}},
  \citenamefont {{Godignon}},\ and\ \citenamefont {{Camassel}}}]{tiberj}%
  \BibitemOpen
  \bibfield  {author} {\bibinfo {author} {\bibfnamefont {A.}~\bibnamefont
  {{Tiberj}}}, \bibinfo {author} {\bibfnamefont {J.~R.}\ \bibnamefont
  {{Huntzinger}}}, \bibinfo {author} {\bibfnamefont {N.}~\bibnamefont
  {{Camara}}}, \bibinfo {author} {\bibfnamefont {P.}~\bibnamefont
  {{Godignon}}}, \ and\ \bibinfo {author} {\bibfnamefont {J.}~\bibnamefont
  {{Camassel}}},\ }\href@noop {} {\bibfield  {journal} {\bibinfo  {journal}
  {ArXiv e-prints}\ } (\bibinfo {year} {2012})},\ \Eprint
  {http://arxiv.org/abs/1212.1196} {arXiv:1212.1196 [cond-mat.mes-hall]}
  \BibitemShut {NoStop}%
\end{thebibliography}
\end{document}